\documentclass[aip,superscriptaddress, amsmath,amssymb, reprint]{revtex4-1}
\usepackage{graphicx}
\usepackage{dcolumn}
\usepackage{bm}
\usepackage{amsmath}
\usepackage{cases}
\usepackage{hyperref}
\usepackage[mathlines]{lineno}
\usepackage[english]{babel}
\usepackage{savesym}
\usepackage{amsmath,amsthm,amssymb}
\usepackage{amsfonts}
\usepackage{xspace}
\savesymbol{setrefcountdefault}
\usepackage{graphicx}

\usepackage[separate-uncertainty=true]{siunitx}
\DeclareSIUnit[number-unit-product = {}]
\decade{\text{dec}}
\usepackage{dcolumn}
\usepackage{lmodern}

\begin{document}

\preprint{AIP/123-QED}

\title{Free-Layer-Thickness-dependence of the Spin Galvanic Effect with Spin Rotation Symmetry}

\author{Wafa S. Aljuaid}
\affiliation{Department of Physics and Astronomy, University of Denver, Denver,Colorado 80208,USA}
\affiliation{Department of Mechanical and Materials Engineering, University of Denver, Denver,Colorado 80208,USA}
\affiliation{Department of Chemistry, Taif University, Taif 26571, Saudi Arabia}


\author{Shane R. Allen}%
\affiliation{Department of Physics and Astronomy, University of Denver, Denver,Colorado 80208,USA}
\author{Angie Davidson}%
\affiliation{Department of Physics and Astronomy, University of Denver, Denver,Colorado 80208,USA}

\author{Xin Fan}
\thanks{Author to whom correspondence should be addressed.}
 \email{xin.fan@du.edu}
\affiliation{Department of Physics and Astronomy, University of Denver, Denver,Colorado 80208,USA}


\date{\today}

\begin{abstract}
Spin-orbit coupling near the surface of a ferromagnetic metal gives rises to spin-to-charge conversion with symmetry different from the conventional inverse spin Hall effect. We have previously observed this spin galvanic effect with spin rotation symmetry (SGE-SR) in a spin valve under a temperature gradient. Here we show there are two processes that contribute to the SGE-SR, one of which is sensitive to the free magnetic layer thickness, while the other only depends on the interface of the free layer. Based on the free-layer-thickness-dependent study, we extrapolate the spin diffusion length of Py to be $3.9 \pm 0.2$\ nm. We also propose that the SGE-SR can help to quantitatively study the spin Seebeck effect in metallic magnetic films.
%
\end{abstract}

\maketitle

%

Spin-charge interconversion enabled by spin-orbit coupling SOC has been intensively studied in nonmagnetic materials (NM) \cite{r1,r2,r26}.It has been shown that an in-plane charge current in a ferromagnetic material (FM)/NM bilayer can generate a spin-orbit torque, which can switch the FM magnetization \cite{r3,r4},drive magnetic auto-oscillations at microwave frequencies \cite{r5,r6}, and move magnetic domain walls and skyrmions \cite{r7,r8,r9}.The spin galvanic effect (SGE) in the FM/NM bilayer  has potential applications in thermo-electric converters \cite{r10}, and terahertz pulse emitters \cite{r11}, where spin currents generated by heat and femtosecond laser pulse, respectively, are converted into electric fields by SOC in the NM. In these two effects, the spin-charge interconversion is often thought to take place in the NM, which is typically a heavy metal such as Pt and Ta.\ The spin-charge interconversion in the NM generally follows symmetry such that the charge current, spin current and spin polarization are all orthogonal with each other \cite{r25}.\ Recently, we have discovered a magnetization-dependent spin-charge interconversion near the surface of a FM,\ in which the spin polarization is transverse to the magnetization \cite{r12}.\ We refer to this magnetization-dependent SGE as the SGE with spin rotation symmetry (SGE-SR), as the effect is rotated with respect to the conventional SGE (SGE-C). In this letter, we investigate the FM thickness-dependence of the SGE-SR in a spin-valve structure.We show that there are two contributions to the measured signals, one of which depends on the spin diffusion length of the FM, and the other is independent of the FM thickness.\ Our findings provide a comprehensive understanding of the heat-driven SGE-SR experiment with metallic ferromagnet and will pave the road for future investigation of spin-orbit effects in magnetically-ordered systems.

\begin{figure*}
	\includegraphics[scale=0.55]{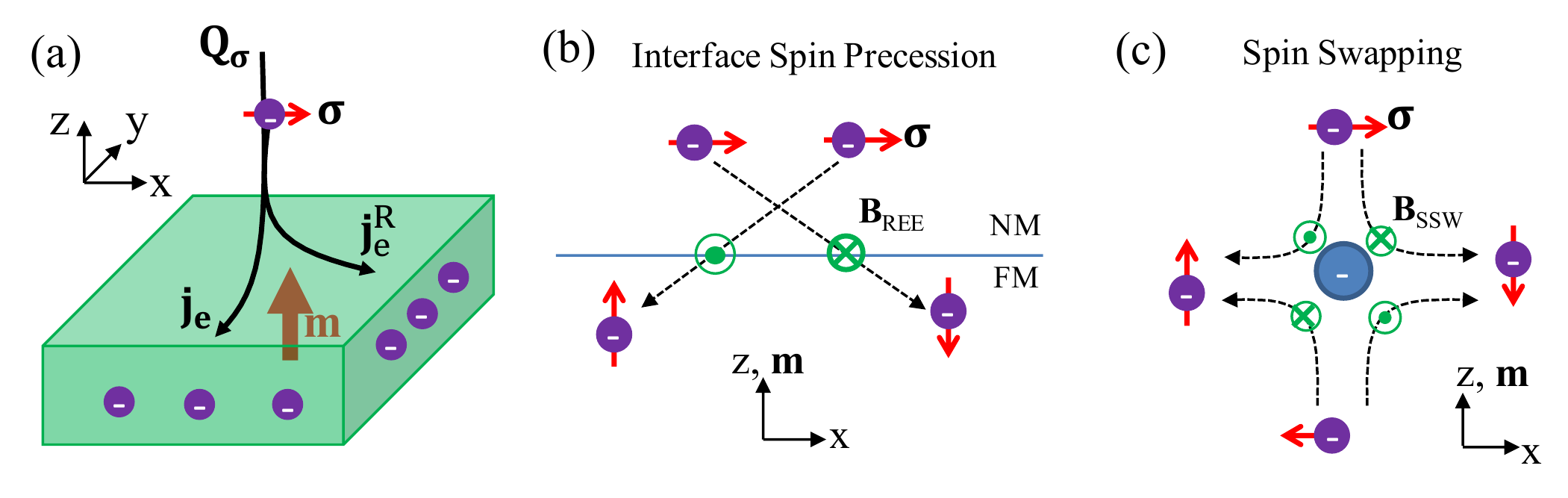}
	\caption{\label{fig1} (a) Illustration of the Spin galvanic effect (SGE) in a ferromagnetic material (FM).(b) Illustration of the interface spin precession mechanism. As electrons cross the NM/FM interface, the spins (red arrows) precess around an effective magnetic field $\mathrm{B_{REE}}$(green vectors), induced by the Rashba-Edelstein spin-orbit coupling.(c) Illustration of the spin swapping mechanism in the FM. When an electron is scattered by an impurity(blue circle), the spins precess around an effective magnetic field $\mathrm{B_{SSW}}$. }
\end{figure*}

The SGE-SR is illustrated in Fig.~\ref{fig1}(a).When a spin current(\textbf{Q$_\sigma$})with spin polarization (\boldsymbol{\mathrm{$\sigma$}})transverse to the magnetization (\textbf{m}) of a FM is injected into the FM, it is usually assumed that the transverse spins undergo fast precession and dephasing due to the exchange interaction. The dephasing quickly transfers angular momentum from transverse spins to the magnetization, which generates the well-known spin transfer torque \cite{r13}. However, this model neglects the effects of SOC. When SOC is taken into account, based on symmetry we expect the spin current to generate an in-plane charge current that can be decomposed into \textbf{j$_e$} with the SGE-C symmetry and \textbf{j$_e^\mathrm{R}$} with the SGE-SR symmetry\cite{r12}.

\begin{equation}
\label{eq1}
	\textbf{j$_e$}  = \frac{2e}{\hbar} \theta \  \textbf{Q$_{\boldsymbol \sigma}$} \times \textbf{$\boldsymbol{\sigma}$}
\end{equation}
\begin{equation}
\label{eq2}
\textbf{j$_e^\mathrm{R}$} = \frac{2e}{\hbar} \theta^R \  \textbf{Q$_{\boldsymbol \sigma}$} \times ( \textbf{${\boldsymbol \sigma}$} \times \boldsymbol {\textbf{m}})
\end{equation}
where $\theta$ and $\theta^R$ are unitless spin-to-charge conversion coefficients for the SGE-C and SGE-SR, respectively, {\textit{e}} is the electron charge and $\hbar$ is the reduced Planck constant.

The generation of the SGE-SR in the FM can be phenomenologically understood by the spin rotation picture. As discussed earlier, transverse spin rotates around magnetization, giving rise to an average spin component in the direction of \boldsymbol{$\sigma $}$\times$$ \textbf{m}$. The SOC acting on this spin component can give rise to the \textbf{j$_e^R$} described by Eq.~\ref{eq2}. 

There are two possible microscopic mechanisms that lead to the SGE-SR: the interface spin precession due to the SOC of the FM/NM interface \cite{r14}; and the spin swapping effect \cite{r15} due to the bulk SOC of the FM. The interface spin precession is illustrated in Fig.\ref{fig1}(b). As an electron crosses the FM/NM interface, it experiences an effective magnetic field due to the Rashba SOC, \textbf{B$_\mathrm{REE}$}$  \propto {\alpha}_\mathrm{R}  \ \textbf{n} \times \textbf{k}$, where  $ {\alpha}_\mathrm{R} $ is the Rashba coefficient, \textbf{k} is the wave vector of the electron and \textbf{n} is the interface normal. The spin of the electron, \boldsymbol{$\sigma$}, can thus precess around \textbf{B$_\mathrm{REE}$} , giving rise to a new component of spin, \textbf{B$_\mathrm{REE}$} $ \times$$\boldsymbol{\sigma}$.

Depending on whether this new spin component is parallel or antiparallel with the magnetization \textbf{m},\ the electron gains different velocities in the FM, which leads to a net electric current flow in the x-direction, consistent with \textbf{j$_e^R$} in Fig.~\ref{fig1}(a).

Alternatively, a transverse spin current flowing in the FM can also produce an electric current with the SGE-SR symmetry due to the spin swapping effect. The spin swapping effect is closely related to skew scattering \cite{r16}. As shown in Fig.~\ref{fig1}(c), when an electron with spin polarization in the x-direction flows in the z-direction, it can be scattered by an impurity center. In the rest frame of the electron, SOC gives rise to a scattering-path-dependent magnetic field, \textbf{B}$_\mathrm{SSW}$.\ The electron spin, which is orthogonal to \textbf{B}$_\mathrm{SSW}$, undergoes precession. The net effect is the generation of a new spin current, whose propagation direction and spin polarization are swapped from the injected spin current. Since the new spin current has polarization parallel or antiparallel with the magnetization, corresponding to majority or minority electrons in the FM, it also yields a net electric current flowing in the x-direction.     

The spin swapping mechanism and the interface spin precession mechanism yield the same spin rotation symmetry, as described by Eq. (2). In both mechanisms, we expect the SGE-SR to occur on a very short length scale due to spin dephasing. Since the spin dephasing length of transverse spins has been found to be less than 1 nm in typical magnetic transition metals \cite{r17}, the SGE-SR can be treated as an interface effect.

\begin{figure*}
	\includegraphics[scale=0.55]{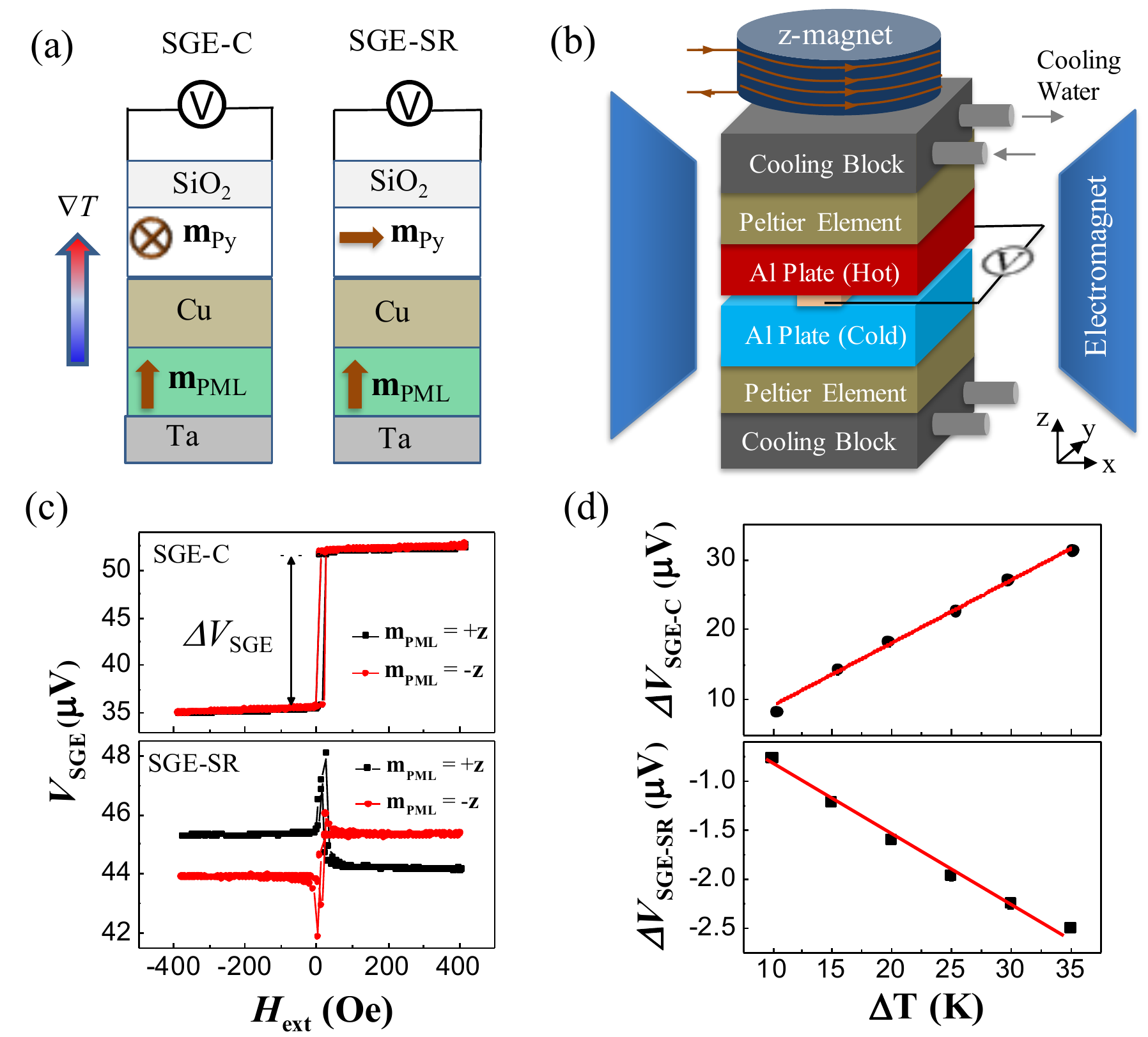}
	\caption{\label{fig2}(a) Two measurement configurations: the voltage detection is perpendicular to the Py magnetization, m$_\mathrm{Py}$, in the SGE-C configuration, while parallel with m$_\mathrm{Py}$ in the SGE-SR configuration.(b)Experimental apparatus (not drawn to scale).\ The sample is  between the two Aluminum plates.\ (c) Exemplary SGE signals measured in the SGE-C and SGE-SR configurations.\ $H_{\mathrm{ext}}$ is the in-plane external magnetization field (d) Temperature dependence of the SGE-C and SGE-SR signals. The red curve is linear fitting to $\Delta$T.}
\end{figure*}

In order to experimentally observe the SGE-SR, we use a spin valve structure consisting of two orthogonally magnetized layers as shown in Fig.\ref{fig2}(a).\ The multilayer film is fabricated by magnetron sputtering with the structure: Ta(3)/PML/Cu(4)/Py(2)/SiO$_2$(4), where PML = Pt(3)/Co(0.6) is a perpendicularly magnetized layer, Py = Ni$_{80}$Fe$_{20}$ and the thicknesses in parenthesis are in nanometers.\ By applying a perpendicular temperature gradient, the spin Seebeck effect \cite{r18} in one of the magnetic layer generates a spin current with spin polarization transverse to the other magnetic layer. An in-plane voltage can be measured in two configurations: the SGE-C configuration, namely the configuration sensitive to the anomalous Nernst effect, the inverse spin Hall effect and other possible SGE with the conventional spin Hall symmetry; and the SGE-SR configuration, where SGE-SR signals are the dominant signals.

The film is deposited on a precut silicon substrate slab with a $1 \mathrm{\mu m}$ thermal oxide surface layer and a lateral size of 2 mm $\times$ 20 mm.\ The measurement setup is shown in Fig.\ref{fig2}(b).\ The sample is placed in between Peltier elements , which are used to generate perpendicular temperature gradient across the sample. Thin aluminum blocks with embedded thermocouples are inserted between the sample and the Peltier elements to monitor the temperature and to enhance the uniformity of heat flow.The sample is attached to the top surface of bottom aluminum plate via a thin piece of Micro-faze thermal sheet. The bottom surface of the top aluminum plate is covered with Sil-pad from Bergquist company and then a thin Kapton tape to establish a uniform thermal contact while avoiding electric contact. The top aluminum plate is pressed against the sample by spring-loaded screws.\ The other sides of the Peltier elements are in contact with cooling blocks with circulating cooling water, which help regulate the temperature stability.\ The whole thermoelectric setup is placed in an electromagnet on a rotatable base, which can apply in-plane magnetic field in an arbitrary direction. There is also a 'z-magnet' on top of the thermoelectric setup, to generate a pulsed magnetic field in the out-of-plane direction, which initializes the magnetization of the PML to $+z$ or $ -z$ directions. 

Exemplary signals are shown in Fig.~\ref{fig2}(c).\ As in-plane magnetic field $H_{\mathrm{ext}}$ sweeps,\ the voltage signal in the SGE-C configuration corresponds to the hysteresis of the Py layer.\ This SGE-C signal is independent to the PML magnetization \textbf{m}$_{\mathrm{PML}}$, as expected from the symmetry of the anomalous Nernst effect and the SGE-C described by Eq.(1).\ In the SGE-SR configuration,the voltage signal reverses sign when \textbf{m}$_{\mathrm{PML}}$ reverses, consistent with Eq~\ref{eq2}.The spikes at low field are due to the ANE because magnetization tilts away from the external magnetic field. It should be noted that even in the SGE-SR configuration, a small misalignment between the magnetic field and the voltage detection direction can result in contributions from the SGE-C signal.\ Therefore,\ we further isolate the SGE-C and SGE-SR signals by adding and subtracting the voltage signals measured when \textbf{m}$_{\mathrm{PML}}$  is polarized in the  $+$z or $ -$z directions.

\begin{equation}
\label{eq3}
\begin{cases} 
\Delta V_\mathrm{{SGE-C}}=\Delta \mathrm{V}_\mathrm{{SGE}} (\mathrm{m}_\mathrm{{PML}} = +z ) \\ + \Delta \mathrm{V}_\mathrm{{SGE}} (\mathrm{m}_\mathrm{{PML}} = -z) \\
\Delta \mathrm{V}_\mathrm{{SGE-SR}} = \Delta \mathrm{V}_\mathrm{{SGE}} (\mathrm{m}_\mathrm{{PML}} = +z ) \\ - \Delta \mathrm{V}_\mathrm{{SGE}} (\mathrm{m}_\mathrm{{PML}} = -z)
\end{cases}
\end{equation}

\begin{figure*}
	\includegraphics[scale=0.55]{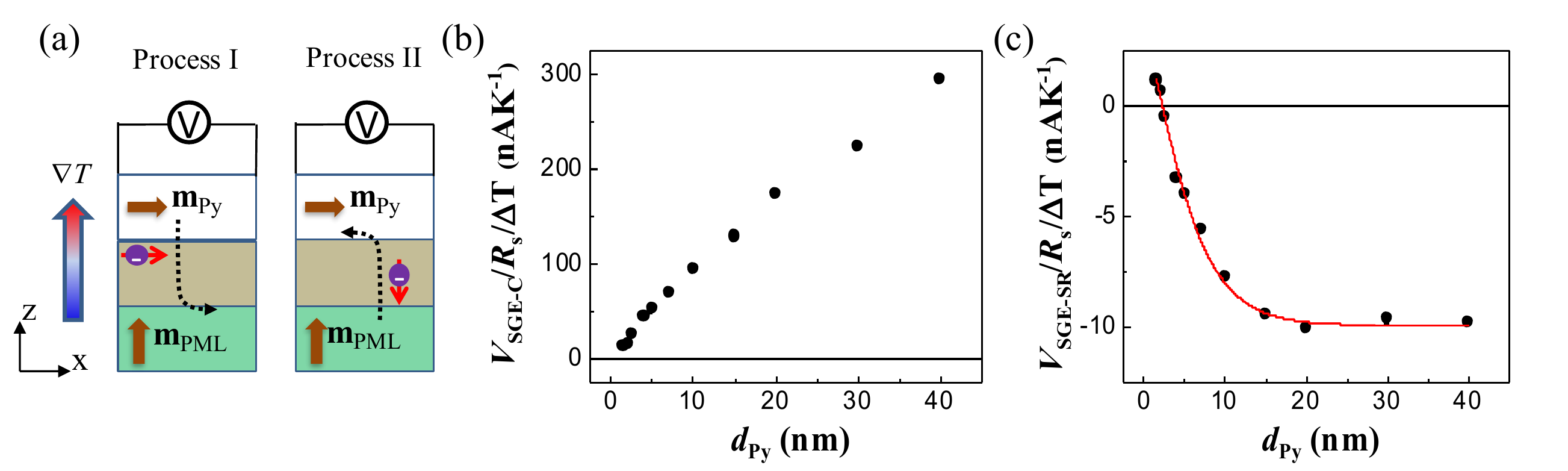}
	\caption{\label{fig3}(a)Illustration of the two processes that give rise to the SGE-SR signal with the boundary conditions used in the simulation of process I.(b,c) Experimentally measured Py-thickness-dependent SGE-C and SGE-SR signals normalized by the sheet resistance $R_s$ and temperature difference $\Delta$T. The  red line is the fitting curve according to Eq.(6).} 
\end{figure*}

where $\Delta V_\mathrm{{SGE}}$ is the intercept difference between the voltage curves measured at positive and negative fields, as labeled in Fig.~\ref{fig2}(c). We varied the temperature difference across the sample. As shown in Fig.~\ref{fig2}(d), both the SGE-C and SGE-SR signals exhibit near linear dependence on the temperature difference ranging from 10 to 35 K.In the rest of the experiments, the temperature difference was kept near 30K, where the temperature of top and bottom Aluminum plates were about $40^{\circ}C$ and $10^{\circ}C$, respectively.The temperature of the films is around room temperature. 

Because both the PML and Py layers are magnetic and possess SOC, the SGE-SR signals should arise from two independent processes, as shown in Fig.~\ref{fig3}(a). In process I, the spin Seebeck effect in Py generates a spin current, which flows toward the PML.\ This spin current is polarized parallel with Py magnetization and hence perpendicular to the PML magnetization. The spin current generates a SGE-SR signal near the PML/Cu interface. Since the spin Seebeck effect in Py can be described by the drift-diffusion model \cite{r19},we expect this part of the SGE-SR signal to have an exponential-like dependence on the Py thickness with the characteristic length being the spin diffusion length in Py. In process II, the spin Seebeck effect in PML generates spin current, which flows toward the Py layer.This spin current is polarized parallel with PML magnetization, hence transverse to the Py magnetization. The spin current generates a SGE-SR signal near the Py/Cu interface. As discussed earlier, the SGE-SR can be treated as an interface effect due to the strong spin dephasing.\ Therefore, the SGE-SR signal due to process II should be nearly independent of the Py thickness.

Here we use the drift-diffusion model to simulate the SGE-SR due to process I.\ The spin Seebeck effect of Py drives a spin current flowing in the z-direction with spin polarization in the x-direction.\ The spin chemical potential $\mu_x$  and spin current $Q_x$ in Py due to the spin Seebeck effect can be expressed as \cite{r24} 

\begin{equation}
\label{eq4}
{}
\begin{cases}
\frac{d^2}{dz^2} \mu_x = \frac{1}{\lambda ^2}  \mu_x  \\ 
Q_x= - \sigma (1 - P^2)  \frac{d}{dz} \mu_x + Q_0\\
\end{cases}       
\end{equation}

where $\lambda$, $\sigma$ and P are the spin diffusion length, electric conductivity and spin polarization of Py, respectively. $Q_0=- \frac{\hbar}{2e}  \sigma (1-P^2) \frac{S_\uparrow - S_\downarrow}{2}  \frac{dT}{dz}$ is the spin current generated by the spin Seebeck effect in an infinitely thick sample, where $S_\uparrow$ and $S_\downarrow$ are the Seebeck coefficients for majority and minority spins, respectively, and \textit{T} is the temperature.

Since the spin current must vanish at the $\mathrm{Py}/\mathrm{SiO_2}$ interface, and the spin current at the PML/Cu interface can be modeled by the magneto-electric circuit theory\cite{rr30}, we calculate the spin current flowing into the PML to be

\begin{equation}
\label{eq5}
Q^\mathrm{PML/Cu}_x = \frac{\mathrm{cosh} \frac{d_{\mathrm {py}}}{\lambda} - 1}{\mathrm{cosh} \frac{d_{\mathrm {py}}}{\lambda} + \frac{\sigma (1- P^2)}{2G ^{\uparrow \downarrow} \lambda} \mathrm{sinh}\frac{d_{\mathrm {py}}}{\lambda}}Q_0
\end{equation}

where ${d_{\mathrm {py}}}$ and\ $G^{\uparrow \downarrow}$ are the Py thickness and the spin mixing conductance at the Co/Cu interface. Therefore,the SGE-SR voltage signal due to process I, $\Delta V^I _\mathrm{{SGE - SR}}$, can be expressed as $\Delta V^I _\mathrm{{SGE - SR}} = 4 \frac{2e}{\hbar} Q^\mathrm{PML/Cu}_x \ \theta^Rl \delta R_s$,  where $\theta^R$  is the coefficient for the SGE-SR at the PML/Cu interface, $l$ is the length of the sample, $\delta$ is the effective thickness of the PML that participates in the SGE-SR, $R_s$ is the square resistance of the entire film stack, and the factor of 4 arises from the definition in Eq.\ref{eq2}.\ The total SGE-SR voltage signal normalized by the sheet resistance $R_s$ can be expressed as

\begin{multline}
\label{eq6}
\frac{\Delta V _\mathrm{{SGE - SR}}}{R_s}   =\frac{1}{R_s}  (\Delta V^{I} _\mathrm{{SGE - SR}} + \Delta V^{II} _\mathrm{{SGE - SR}}) \\ = \frac{4 \ \mathrm{cosh} \frac{d_{\mathrm {py}}}{\lambda} - 4}{\mathrm{cosh} \frac{d_{\mathrm {py}}}{\lambda} + \frac{\sigma  (1- P^2)}{2G ^{\uparrow \downarrow} \lambda} \mathrm{sinh}\frac{d_{\mathrm {py}}}{\lambda}}   
\frac{2e}{\hbar} Q_0 \theta^R l\delta  + \frac{\Delta V^{II}  _\mathrm{{SGE - SR}}}{R_s}
\end{multline}

We experimentally measured the SGE-C and SGE-SR voltage signals with various ${d_{\mathrm {py}}}$.The voltage signals normalized by the square resistance and temperature are plotted against ${d_{\mathrm {py}}}$, as shown in Fig.~\ref{fig3}(b) and (c). The SGE-C signal exhibits monotonic increase with \textit d.This is because a large portion of the SGE-C comes from the anomalous Nernst effect, which contributes more significantly as Py gets thicker. The SGE-SR signal starts at a positive value when Py is thin.\ As Py gets thicker, the SGE-SR signal crosses zero and eventually saturates at a negative value.\ Such a behavior can be explained by a unique case of Eq.~\ref{eq6},where process I ($\Delta V^{I} _{\mathrm{SGE - SR}}$) contributes negative value and process II ($\Delta V^{II} _{\mathrm{SGE - SR}}$) contributes positive value to the measured SGE-SR signal.\ When Py is thin, $\Delta V^{II} _{\mathrm{SGE - SR}}$ is greater than $\Delta V^{I} _{\mathrm{SGE - SR}}$.\ As Py gets thicker, $\Delta V^{I} _{\mathrm{SGE - SR}}$ increases and eventually overwhelms $\Delta V^{II} _{\mathrm{SGE - SR}}$, which is nearly independent of the Py thickness.\ Using Eq.~\ref{eq6} to fit Fig.~\ref{fig3}(c) with the Levenberg-Marquardt algorithm, we obtained the spin diffusion length in Py to be $3.9 \pm 0.2\ \mathrm{nm}$,\ consistent with previous reported value\cite{r29} , and  $\frac{\Delta V^{II} _\mathrm{{SGE - SR}}}{\mathrm{R_s}}$ = $3.0 \pm 0.3  \ \mathrm{nA}$ .\ Here the parameters used are $\sigma= 4.8 \times 10 ^{6} \Omega^{-1} \mathrm{m^{-1}}$ (extrapolated from Py-thickness-dependent resistance measurement ), $G^{\uparrow \downarrow}=0.4$ $\times 10 ^{15}  \Omega^{-1}  \mathrm{m^{-2}} ${(Ref. \citenum{r21})}, $P=50\% $ {(Ref. \citenum{r27})}.

Besides the spin diffusion length of Py, fitting to Eq.~\ref{eq6} can also give rise to $Q_0 \theta^R \delta$ to be ($-4.5\pm 0.3) \times 10 ^{-5}\mathrm{\frac {\hbar}{2e} A/m}$ for $\Delta$T= 30 K, where $Q_0$ relates to the spin Seebeck effect of the Py, and $\theta^R \delta$ describes the SGE-SR at the PML/Cu interface.\ While the temperature gradient can be consistently generated in our measurement, it is difficult to quantify the exact temperature gradient across the film due to unknown thermal contact resistance at various interfaces \cite{r31}.In our SGE-SR measurement, the temperature gradient $ \nabla $T can be calibrated by the SGE-C signals in samples with thick Py, which is dominated by the ANE of Py : $\nabla T=   \frac{V_{\mathrm{SGE-C}}}{l{\alpha_{N}}}  \frac{R_\mathrm{s}}{R_{\mathrm{s,py}}}$  , where $R_{\mathrm{s,py}}$ is the calculated sheet resistance of Py based on the Py conductivity, $\alpha_{N}$ is the ANE coefficient of Py.\ If  $\theta^R \delta$ and $\alpha_{N}$ can be extrapolated from other techniques\cite{r12,r32}, this study can provide a unique tool to quantify the spin Seebeck effect in a metallic ferromagnetic material . Although longitudinal spin transport has been widely used to study the spin Seebeck effect from ferromagnetic insulators \cite{r22}, it is challenging to measure the spin Seebeck effect in metallic FM/NM bilayer with the same technique, due to the complication from the anomalous Nernst effect of the FM\cite{r23} . The measurement of the spin Seebeck effect in metallic FM has been realized by a lateral nonlocal spin transport technique \cite{r24}, which requires extensive nanolithography. In our measurement of the SGE-SR, no lithography is needed, and complication due to the anomalous Nernst effect is removed due to the unique symmetry.

In conclusion,we have demonstrated that the SGE-SR in a metallic spin valve is composed of two contributions, which can be separated by FM-thickness-dependent study.A spin diffusion length of $3.9 \pm 0.2 \mathrm{nm}$  for Py is extrapolated.\ The SGE-SR may potentially be used for quantitative measurement of the spin Seebeck effect if the SGE-SR coefficients can be determined by an independent technique.\\

This work is supported by the National Science Foundation under Grant Number ECCS-1738679, and by the DU PROF fund. W.S.A Would like to acknowledge support from Saudi Arabia Cultural Mission .


\bibliography{aipsamp}

\begin{thebibliography}{30}%
\makeatletter
\providecommand \@ifxundefined [1]{%
 \@ifx{#1\undefined}
}%
\providecommand \@ifnum [1]{%
 \ifnum #1\expandafter \@firstoftwo
 \else \expandafter \@secondoftwo
 \fi
}%
\providecommand \@ifx [1]{%
 \ifx #1\expandafter \@firstoftwo
 \else \expandafter \@secondoftwo
 \fi
}%
\providecommand \natexlab [1]{#1}%
\providecommand \enquote  [1]{``#1''}%
\providecommand \bibnamefont  [1]{#1}%
\providecommand \bibfnamefont [1]{#1}%
\providecommand \citenamefont [1]{#1}%
\providecommand \href@noop [0]{\@secondoftwo}%
\providecommand \href [0]{\begingroup \@sanitize@url \@href}%
\providecommand \@href[1]{\@@startlink{#1}\@@href}%
\providecommand \@@href[1]{\endgroup#1\@@endlink}%
\providecommand \@sanitize@url [0]{\catcode `\\12\catcode `\$12\catcode
  `\&12\catcode `\#12\catcode `\^12\catcode `\_12\catcode `\%12\relax}%
\providecommand \@@startlink[1]{}%
\providecommand \@@endlink[0]{}%
\providecommand \url  [0]{\begingroup\@sanitize@url \@url }%
\providecommand \@url [1]{\endgroup\@href {#1}{\urlprefix }}%
\providecommand \urlprefix  [0]{URL }%
\providecommand \Eprint [0]{\href }%
\providecommand \doibase [0]{http://dx.doi.org/}%
\providecommand \selectlanguage [0]{\@gobble}%
\providecommand \bibinfo  [0]{\@secondoftwo}%
\providecommand \bibfield  [0]{\@secondoftwo}%
\providecommand \translation [1]{[#1]}%
\providecommand \BibitemOpen [0]{}%
\providecommand \bibitemStop [0]{}%
\providecommand \bibitemNoStop [0]{.\EOS\space}%
\providecommand \EOS [0]{\spacefactor3000\relax}%
\providecommand \BibitemShut  [1]{\csname bibitem#1\endcsname}%
\let\auto@bib@innerbib\@empty
\bibitem [{\citenamefont {Saitoh}\ \emph {et~al.}(2006)\citenamefont {Saitoh},
  \citenamefont {Ueda}, \citenamefont {Miyajima},\ and\ \citenamefont
  {Tatara}}]{r1}%
  \BibitemOpen
  \bibfield  {author} {\bibinfo {author} {\bibfnamefont {E.}~\bibnamefont
  {Saitoh}}, \bibinfo {author} {\bibfnamefont {M.}~\bibnamefont {Ueda}},
  \bibinfo {author} {\bibfnamefont {H.}~\bibnamefont {Miyajima}}, \ and\
  \bibinfo {author} {\bibfnamefont {G.}~\bibnamefont {Tatara}},\ }\bibfield
  {title} {\enquote {\bibinfo {title} {Conversion of spin current into charge
  current at room temperature: Inverse spin-hall effect},}\ }\href@noop {}
  {\bibfield  {journal} {\bibinfo  {journal} {Applied physics letters}\
  }\textbf {\bibinfo {volume} {88}},\ \bibinfo {pages} {182509} (\bibinfo
  {year} {2006})}\BibitemShut {NoStop}%
\bibitem [{\citenamefont {Kimura}\ \emph {et~al.}(2007)\citenamefont {Kimura},
  \citenamefont {Otani}, \citenamefont {Sato}, \citenamefont {Takahashi},\ and\
  \citenamefont {Maekawa}}]{r2}%
  \BibitemOpen
  \bibfield  {author} {\bibinfo {author} {\bibfnamefont {T.}~\bibnamefont
  {Kimura}}, \bibinfo {author} {\bibfnamefont {Y.}~\bibnamefont {Otani}},
  \bibinfo {author} {\bibfnamefont {T.}~\bibnamefont {Sato}}, \bibinfo {author}
  {\bibfnamefont {S.}~\bibnamefont {Takahashi}}, \ and\ \bibinfo {author}
  {\bibfnamefont {S.}~\bibnamefont {Maekawa}},\ }\bibfield  {title} {\enquote
  {\bibinfo {title} {Room-temperature reversible spin hall effect},}\
  }\href@noop {} {\bibfield  {journal} {\bibinfo  {journal} {Physical review
  letters}\ }\textbf {\bibinfo {volume} {98}},\ \bibinfo {pages} {156601}
  (\bibinfo {year} {2007})}\BibitemShut {NoStop}%
\bibitem [{\citenamefont {Hoffmann}(2013)}]{r26}%
  \BibitemOpen
  \bibfield  {author} {\bibinfo {author} {\bibfnamefont {A.}~\bibnamefont
  {Hoffmann}},\ }\bibfield  {title} {\enquote {\bibinfo {title} {Spin hall
  effects in metals},}\ }\href@noop {} {\bibfield  {journal} {\bibinfo
  {journal} {IEEE transactions on magnetics}\ }\textbf {\bibinfo {volume}
  {49}},\ \bibinfo {pages} {5172--5193} (\bibinfo {year} {2013})}\BibitemShut
  {NoStop}%
\bibitem [{\citenamefont {Liu}\ \emph {et~al.}(2012{\natexlab{a}})\citenamefont
  {Liu}, \citenamefont {Pai}, \citenamefont {Li}, \citenamefont {Tseng},
  \citenamefont {Ralph},\ and\ \citenamefont {Buhrman}}]{r3}%
  \BibitemOpen
  \bibfield  {author} {\bibinfo {author} {\bibfnamefont {L.}~\bibnamefont
  {Liu}}, \bibinfo {author} {\bibfnamefont {C.-F.}\ \bibnamefont {Pai}},
  \bibinfo {author} {\bibfnamefont {Y.}~\bibnamefont {Li}}, \bibinfo {author}
  {\bibfnamefont {H.}~\bibnamefont {Tseng}}, \bibinfo {author} {\bibfnamefont
  {D.}~\bibnamefont {Ralph}}, \ and\ \bibinfo {author} {\bibfnamefont
  {R.}~\bibnamefont {Buhrman}},\ }\bibfield  {title} {\enquote {\bibinfo
  {title} {Spin-torque switching with the giant spin hall effect of
  tantalum},}\ }\href@noop {} {\bibfield  {journal} {\bibinfo  {journal}
  {Science}\ }\textbf {\bibinfo {volume} {336}},\ \bibinfo {pages} {555--558}
  (\bibinfo {year} {2012}{\natexlab{a}})}\BibitemShut {NoStop}%
\bibitem [{\citenamefont {Miron}\ \emph {et~al.}(2011)\citenamefont {Miron},
  \citenamefont {Garello}, \citenamefont {Gaudin}, \citenamefont {Zermatten},
  \citenamefont {Costache}, \citenamefont {Auffret}, \citenamefont {Bandiera},
  \citenamefont {Rodmacq}, \citenamefont {Schuhl},\ and\ \citenamefont
  {Gambardella}}]{r4}%
  \BibitemOpen
  \bibfield  {author} {\bibinfo {author} {\bibfnamefont {I.~M.}\ \bibnamefont
  {Miron}}, \bibinfo {author} {\bibfnamefont {K.}~\bibnamefont {Garello}},
  \bibinfo {author} {\bibfnamefont {G.}~\bibnamefont {Gaudin}}, \bibinfo
  {author} {\bibfnamefont {P.-J.}\ \bibnamefont {Zermatten}}, \bibinfo {author}
  {\bibfnamefont {M.~V.}\ \bibnamefont {Costache}}, \bibinfo {author}
  {\bibfnamefont {S.}~\bibnamefont {Auffret}}, \bibinfo {author} {\bibfnamefont
  {S.}~\bibnamefont {Bandiera}}, \bibinfo {author} {\bibfnamefont
  {B.}~\bibnamefont {Rodmacq}}, \bibinfo {author} {\bibfnamefont
  {A.}~\bibnamefont {Schuhl}}, \ and\ \bibinfo {author} {\bibfnamefont
  {P.}~\bibnamefont {Gambardella}},\ }\bibfield  {title} {\enquote {\bibinfo
  {title} {Perpendicular switching of a single ferromagnetic layer induced by
  in-plane current injection},}\ }\href@noop {} {\bibfield  {journal} {\bibinfo
   {journal} {Nature}\ }\textbf {\bibinfo {volume} {476}},\ \bibinfo {pages}
  {189} (\bibinfo {year} {2011})}\BibitemShut {NoStop}%
\bibitem [{\citenamefont {Liu}\ \emph {et~al.}(2012{\natexlab{b}})\citenamefont
  {Liu}, \citenamefont {Pai}, \citenamefont {Ralph},\ and\ \citenamefont
  {Buhrman}}]{r5}%
  \BibitemOpen
  \bibfield  {author} {\bibinfo {author} {\bibfnamefont {L.}~\bibnamefont
  {Liu}}, \bibinfo {author} {\bibfnamefont {C.-F.}\ \bibnamefont {Pai}},
  \bibinfo {author} {\bibfnamefont {D.}~\bibnamefont {Ralph}}, \ and\ \bibinfo
  {author} {\bibfnamefont {R.}~\bibnamefont {Buhrman}},\ }\bibfield  {title}
  {\enquote {\bibinfo {title} {Magnetic oscillations driven by the spin hall
  effect in 3-terminal magnetic tunnel junction devices},}\ }\href@noop {}
  {\bibfield  {journal} {\bibinfo  {journal} {Physical review letters}\
  }\textbf {\bibinfo {volume} {109}},\ \bibinfo {pages} {186602} (\bibinfo
  {year} {2012}{\natexlab{b}})}\BibitemShut {NoStop}%
\bibitem [{\citenamefont {Liu}, \citenamefont {Lim},\ and\ \citenamefont
  {Urazhdin}(2013)}]{r6}%
  \BibitemOpen
  \bibfield  {author} {\bibinfo {author} {\bibfnamefont {R.}~\bibnamefont
  {Liu}}, \bibinfo {author} {\bibfnamefont {W.}~\bibnamefont {Lim}}, \ and\
  \bibinfo {author} {\bibfnamefont {S.}~\bibnamefont {Urazhdin}},\ }\bibfield
  {title} {\enquote {\bibinfo {title} {Spectral characteristics of the
  microwave emission by the spin hall nano-oscillator},}\ }\href@noop {}
  {\bibfield  {journal} {\bibinfo  {journal} {Physical review letters}\
  }\textbf {\bibinfo {volume} {110}},\ \bibinfo {pages} {147601} (\bibinfo
  {year} {2013})}\BibitemShut {NoStop}%
\bibitem [{\citenamefont {Emori}\ \emph {et~al.}(2013)\citenamefont {Emori},
  \citenamefont {Bauer}, \citenamefont {Ahn}, \citenamefont {Martinez},\ and\
  \citenamefont {Beach}}]{r7}%
  \BibitemOpen
  \bibfield  {author} {\bibinfo {author} {\bibfnamefont {S.}~\bibnamefont
  {Emori}}, \bibinfo {author} {\bibfnamefont {U.}~\bibnamefont {Bauer}},
  \bibinfo {author} {\bibfnamefont {S.-M.}\ \bibnamefont {Ahn}}, \bibinfo
  {author} {\bibfnamefont {E.}~\bibnamefont {Martinez}}, \ and\ \bibinfo
  {author} {\bibfnamefont {G.~S.}\ \bibnamefont {Beach}},\ }\bibfield  {title}
  {\enquote {\bibinfo {title} {Current-driven dynamics of chiral ferromagnetic
  domain walls},}\ }\href@noop {} {\bibfield  {journal} {\bibinfo  {journal}
  {Nature materials}\ }\textbf {\bibinfo {volume} {12}},\ \bibinfo {pages}
  {611} (\bibinfo {year} {2013})}\BibitemShut {NoStop}%
\bibitem [{\citenamefont {Ryu}\ \emph {et~al.}(2013)\citenamefont {Ryu},
  \citenamefont {Thomas}, \citenamefont {Yang},\ and\ \citenamefont
  {Parkin}}]{r8}%
  \BibitemOpen
  \bibfield  {author} {\bibinfo {author} {\bibfnamefont {K.-S.}\ \bibnamefont
  {Ryu}}, \bibinfo {author} {\bibfnamefont {L.}~\bibnamefont {Thomas}},
  \bibinfo {author} {\bibfnamefont {S.-H.}\ \bibnamefont {Yang}}, \ and\
  \bibinfo {author} {\bibfnamefont {S.}~\bibnamefont {Parkin}},\ }\bibfield
  {title} {\enquote {\bibinfo {title} {Chiral spin torque at magnetic domain
  walls},}\ }\href@noop {} {\bibfield  {journal} {\bibinfo  {journal} {Nature
  nanotechnology}\ }\textbf {\bibinfo {volume} {8}},\ \bibinfo {pages} {527}
  (\bibinfo {year} {2013})}\BibitemShut {NoStop}%
\bibitem [{\citenamefont {Jiang}\ \emph {et~al.}(2015)\citenamefont {Jiang},
  \citenamefont {Upadhyaya}, \citenamefont {Zhang}, \citenamefont {Yu},
  \citenamefont {Jungfleisch}, \citenamefont {Fradin}, \citenamefont {Pearson},
  \citenamefont {Tserkovnyak}, \citenamefont {Wang}, \citenamefont {Heinonen}
  \emph {et~al.}}]{r9}%
  \BibitemOpen
  \bibfield  {author} {\bibinfo {author} {\bibfnamefont {W.}~\bibnamefont
  {Jiang}}, \bibinfo {author} {\bibfnamefont {P.}~\bibnamefont {Upadhyaya}},
  \bibinfo {author} {\bibfnamefont {W.}~\bibnamefont {Zhang}}, \bibinfo
  {author} {\bibfnamefont {G.}~\bibnamefont {Yu}}, \bibinfo {author}
  {\bibfnamefont {M.~B.}\ \bibnamefont {Jungfleisch}}, \bibinfo {author}
  {\bibfnamefont {F.~Y.}\ \bibnamefont {Fradin}}, \bibinfo {author}
  {\bibfnamefont {J.~E.}\ \bibnamefont {Pearson}}, \bibinfo {author}
  {\bibfnamefont {Y.}~\bibnamefont {Tserkovnyak}}, \bibinfo {author}
  {\bibfnamefont {K.~L.}\ \bibnamefont {Wang}}, \bibinfo {author}
  {\bibfnamefont {O.}~\bibnamefont {Heinonen}},  \emph {et~al.},\ }\bibfield
  {title} {\enquote {\bibinfo {title} {Blowing magnetic skyrmion bubbles},}\
  }\href@noop {} {\bibfield  {journal} {\bibinfo  {journal} {Science}\ }\textbf
  {\bibinfo {volume} {349}},\ \bibinfo {pages} {283--286} (\bibinfo {year}
  {2015})}\BibitemShut {NoStop}%
\bibitem [{\citenamefont {Kirihara}\ \emph {et~al.}(2012)\citenamefont
  {Kirihara}, \citenamefont {Uchida}, \citenamefont {Kajiwara}, \citenamefont
  {Ishida}, \citenamefont {Nakamura}, \citenamefont {Manako}, \citenamefont
  {Saitoh},\ and\ \citenamefont {Yorozu}}]{r10}%
  \BibitemOpen
  \bibfield  {author} {\bibinfo {author} {\bibfnamefont {A.}~\bibnamefont
  {Kirihara}}, \bibinfo {author} {\bibfnamefont {K.-i.}\ \bibnamefont
  {Uchida}}, \bibinfo {author} {\bibfnamefont {Y.}~\bibnamefont {Kajiwara}},
  \bibinfo {author} {\bibfnamefont {M.}~\bibnamefont {Ishida}}, \bibinfo
  {author} {\bibfnamefont {Y.}~\bibnamefont {Nakamura}}, \bibinfo {author}
  {\bibfnamefont {T.}~\bibnamefont {Manako}}, \bibinfo {author} {\bibfnamefont
  {E.}~\bibnamefont {Saitoh}}, \ and\ \bibinfo {author} {\bibfnamefont
  {S.}~\bibnamefont {Yorozu}},\ }\bibfield  {title} {\enquote {\bibinfo {title}
  {Spin-current-driven thermoelectric coating},}\ }\href@noop {} {\bibfield
  {journal} {\bibinfo  {journal} {Nature materials}\ }\textbf {\bibinfo
  {volume} {11}},\ \bibinfo {pages} {686} (\bibinfo {year} {2012})}\BibitemShut
  {NoStop}%
\bibitem [{\citenamefont {Seifert}\ \emph {et~al.}(2016)\citenamefont
  {Seifert}, \citenamefont {Jaiswal}, \citenamefont {Martens}, \citenamefont
  {Hannegan}, \citenamefont {Braun}, \citenamefont {Maldonado}, \citenamefont
  {Freimuth}, \citenamefont {Kronenberg}, \citenamefont {Henrizi},
  \citenamefont {Radu} \emph {et~al.}}]{r11}%
  \BibitemOpen
  \bibfield  {author} {\bibinfo {author} {\bibfnamefont {T.}~\bibnamefont
  {Seifert}}, \bibinfo {author} {\bibfnamefont {S.}~\bibnamefont {Jaiswal}},
  \bibinfo {author} {\bibfnamefont {U.}~\bibnamefont {Martens}}, \bibinfo
  {author} {\bibfnamefont {J.}~\bibnamefont {Hannegan}}, \bibinfo {author}
  {\bibfnamefont {L.}~\bibnamefont {Braun}}, \bibinfo {author} {\bibfnamefont
  {P.}~\bibnamefont {Maldonado}}, \bibinfo {author} {\bibfnamefont
  {F.}~\bibnamefont {Freimuth}}, \bibinfo {author} {\bibfnamefont
  {A.}~\bibnamefont {Kronenberg}}, \bibinfo {author} {\bibfnamefont
  {J.}~\bibnamefont {Henrizi}}, \bibinfo {author} {\bibfnamefont
  {I.}~\bibnamefont {Radu}},  \emph {et~al.},\ }\bibfield  {title} {\enquote
  {\bibinfo {title} {Efficient metallic spintronic emitters of ultrabroadband
  terahertz radiation},}\ }\href@noop {} {\bibfield  {journal} {\bibinfo
  {journal} {Nature photonics}\ }\textbf {\bibinfo {volume} {10}},\ \bibinfo
  {pages} {483--488} (\bibinfo {year} {2016})}\BibitemShut {NoStop}%
\bibitem [{\citenamefont {Hirsch}(1999)}]{r25}%
  \BibitemOpen
  \bibfield  {author} {\bibinfo {author} {\bibfnamefont {J.}~\bibnamefont
  {Hirsch}},\ }\bibfield  {title} {\enquote {\bibinfo {title} {Spin hall
  effect},}\ }\href@noop {} {\bibfield  {journal} {\bibinfo  {journal}
  {Physical Review Letters}\ }\textbf {\bibinfo {volume} {83}},\ \bibinfo
  {pages} {1834} (\bibinfo {year} {1999})}\BibitemShut {NoStop}%
\bibitem [{\citenamefont {Humphries}\ \emph {et~al.}(2017)\citenamefont
  {Humphries}, \citenamefont {Wang}, \citenamefont {Edwards}, \citenamefont
  {Allen}, \citenamefont {Shaw}, \citenamefont {Nembach}, \citenamefont {Xiao},
  \citenamefont {Silva},\ and\ \citenamefont {Fan}}]{r12}%
  \BibitemOpen
  \bibfield  {author} {\bibinfo {author} {\bibfnamefont {A.~M.}\ \bibnamefont
  {Humphries}}, \bibinfo {author} {\bibfnamefont {T.}~\bibnamefont {Wang}},
  \bibinfo {author} {\bibfnamefont {E.~R.}\ \bibnamefont {Edwards}}, \bibinfo
  {author} {\bibfnamefont {S.~R.}\ \bibnamefont {Allen}}, \bibinfo {author}
  {\bibfnamefont {J.~M.}\ \bibnamefont {Shaw}}, \bibinfo {author}
  {\bibfnamefont {H.~T.}\ \bibnamefont {Nembach}}, \bibinfo {author}
  {\bibfnamefont {J.~Q.}\ \bibnamefont {Xiao}}, \bibinfo {author}
  {\bibfnamefont {T.}~\bibnamefont {Silva}}, \ and\ \bibinfo {author}
  {\bibfnamefont {X.}~\bibnamefont {Fan}},\ }\bibfield  {title} {\enquote
  {\bibinfo {title} {Observation of spin-orbit effects with spin rotation
  symmetry},}\ }\href@noop {} {\bibfield  {journal} {\bibinfo  {journal}
  {Nature communications}\ }\textbf {\bibinfo {volume} {8}},\ \bibinfo {pages}
  {911} (\bibinfo {year} {2017})}\BibitemShut {NoStop}%
\bibitem [{\citenamefont {Stiles}\ and\ \citenamefont {Zangwill}(2002)}]{r13}%
  \BibitemOpen
  \bibfield  {author} {\bibinfo {author} {\bibfnamefont {M.~D.}\ \bibnamefont
  {Stiles}}\ and\ \bibinfo {author} {\bibfnamefont {A.}~\bibnamefont
  {Zangwill}},\ }\bibfield  {title} {\enquote {\bibinfo {title} {Anatomy of
  spin-transfer torque},}\ }\href@noop {} {\bibfield  {journal} {\bibinfo
  {journal} {Physical Review B}\ }\textbf {\bibinfo {volume} {66}},\ \bibinfo
  {pages} {014407} (\bibinfo {year} {2002})}\BibitemShut {NoStop}%
\bibitem [{\citenamefont {Seung-heon}\ \emph {et~al.}(2018)\citenamefont
  {Seung-heon}, \citenamefont {Amin}, \citenamefont {Oh}, \citenamefont {Go},
  \citenamefont {Lee}, \citenamefont {Lee}, \citenamefont {Kim}, \citenamefont
  {Stiles}, \citenamefont {Park},\ and\ \citenamefont {Lee}}]{r14}%
  \BibitemOpen
  \bibfield  {author} {\bibinfo {author} {\bibfnamefont {C.~B.}\ \bibnamefont
  {Seung-heon}}, \bibinfo {author} {\bibfnamefont {V.~P.}\ \bibnamefont
  {Amin}}, \bibinfo {author} {\bibfnamefont {Y.-W.}\ \bibnamefont {Oh}},
  \bibinfo {author} {\bibfnamefont {G.}~\bibnamefont {Go}}, \bibinfo {author}
  {\bibfnamefont {S.-J.}\ \bibnamefont {Lee}}, \bibinfo {author} {\bibfnamefont
  {G.-H.}\ \bibnamefont {Lee}}, \bibinfo {author} {\bibfnamefont {K.-J.}\
  \bibnamefont {Kim}}, \bibinfo {author} {\bibfnamefont {M.~D.}\ \bibnamefont
  {Stiles}}, \bibinfo {author} {\bibfnamefont {B.-G.}\ \bibnamefont {Park}}, \
  and\ \bibinfo {author} {\bibfnamefont {K.-J.}\ \bibnamefont {Lee}},\
  }\bibfield  {title} {\enquote {\bibinfo {title} {Spin currents and
  spin--orbit torques in ferromagnetic trilayers},}\ }\href@noop {} {\bibfield
  {journal} {\bibinfo  {journal} {Nature materials}\ }\textbf {\bibinfo
  {volume} {17}},\ \bibinfo {pages} {509} (\bibinfo {year} {2018})}\BibitemShut
  {NoStop}%
\bibitem [{\citenamefont {Lifshits}\ and\ \citenamefont
  {Dyakonov}(2009)}]{r15}%
  \BibitemOpen
  \bibfield  {author} {\bibinfo {author} {\bibfnamefont {M.~B.}\ \bibnamefont
  {Lifshits}}\ and\ \bibinfo {author} {\bibfnamefont {M.~I.}\ \bibnamefont
  {Dyakonov}},\ }\bibfield  {title} {\enquote {\bibinfo {title} {Swapping spin
  currents: Interchanging spin and flow directions},}\ }\href@noop {}
  {\bibfield  {journal} {\bibinfo  {journal} {Physical review letters}\
  }\textbf {\bibinfo {volume} {103}},\ \bibinfo {pages} {186601} (\bibinfo
  {year} {2009})}\BibitemShut {NoStop}%
\bibitem [{\citenamefont {Nagaosa}\ \emph {et~al.}(2010)\citenamefont
  {Nagaosa}, \citenamefont {Sinova}, \citenamefont {Onoda}, \citenamefont
  {MacDonald},\ and\ \citenamefont {Ong}}]{r16}%
  \BibitemOpen
  \bibfield  {author} {\bibinfo {author} {\bibfnamefont {N.}~\bibnamefont
  {Nagaosa}}, \bibinfo {author} {\bibfnamefont {J.}~\bibnamefont {Sinova}},
  \bibinfo {author} {\bibfnamefont {S.}~\bibnamefont {Onoda}}, \bibinfo
  {author} {\bibfnamefont {A.}~\bibnamefont {MacDonald}}, \ and\ \bibinfo
  {author} {\bibfnamefont {N.}~\bibnamefont {Ong}},\ }\bibfield  {title}
  {\enquote {\bibinfo {title} {Anomalous hall effect},}\ }\href@noop {}
  {\bibfield  {journal} {\bibinfo  {journal} {Reviews of modern physics}\
  }\textbf {\bibinfo {volume} {82}},\ \bibinfo {pages} {1539} (\bibinfo {year}
  {2010})}\BibitemShut {NoStop}%
\bibitem [{\citenamefont {Ghosh}\ \emph {et~al.}(2012)\citenamefont {Ghosh},
  \citenamefont {Auffret}, \citenamefont {Ebels},\ and\ \citenamefont
  {Bailey}}]{r17}%
  \BibitemOpen
  \bibfield  {author} {\bibinfo {author} {\bibfnamefont {A.}~\bibnamefont
  {Ghosh}}, \bibinfo {author} {\bibfnamefont {S.}~\bibnamefont {Auffret}},
  \bibinfo {author} {\bibfnamefont {U.}~\bibnamefont {Ebels}}, \ and\ \bibinfo
  {author} {\bibfnamefont {W.}~\bibnamefont {Bailey}},\ }\bibfield  {title}
  {\enquote {\bibinfo {title} {Penetration depth of transverse spin current in
  ultrathin ferromagnets},}\ }\href@noop {} {\bibfield  {journal} {\bibinfo
  {journal} {Physical review letters}\ }\textbf {\bibinfo {volume} {109}},\
  \bibinfo {pages} {127202} (\bibinfo {year} {2012})}\BibitemShut {NoStop}%
\bibitem [{\citenamefont {Uchida}\ \emph {et~al.}(2008)\citenamefont {Uchida},
  \citenamefont {Takahashi}, \citenamefont {Harii}, \citenamefont {Ieda},
  \citenamefont {Koshibae}, \citenamefont {Ando}, \citenamefont {Maekawa},\
  and\ \citenamefont {Saitoh}}]{r18}%
  \BibitemOpen
  \bibfield  {author} {\bibinfo {author} {\bibfnamefont {K.}~\bibnamefont
  {Uchida}}, \bibinfo {author} {\bibfnamefont {S.}~\bibnamefont {Takahashi}},
  \bibinfo {author} {\bibfnamefont {K.}~\bibnamefont {Harii}}, \bibinfo
  {author} {\bibfnamefont {J.}~\bibnamefont {Ieda}}, \bibinfo {author}
  {\bibfnamefont {W.}~\bibnamefont {Koshibae}}, \bibinfo {author}
  {\bibfnamefont {K.}~\bibnamefont {Ando}}, \bibinfo {author} {\bibfnamefont
  {S.}~\bibnamefont {Maekawa}}, \ and\ \bibinfo {author} {\bibfnamefont
  {E.}~\bibnamefont {Saitoh}},\ }\bibfield  {title} {\enquote {\bibinfo {title}
  {Observation of the spin seebeck effect},}\ }\href@noop {} {\bibfield
  {journal} {\bibinfo  {journal} {Nature}\ }\textbf {\bibinfo {volume} {455}},\
  \bibinfo {pages} {778} (\bibinfo {year} {2008})}\BibitemShut {NoStop}%
\bibitem [{\citenamefont {Holanda}\ \emph {et~al.}(2017)\citenamefont
  {Holanda}, \citenamefont {Santos}, \citenamefont {Cunha}, \citenamefont
  {Mendes}, \citenamefont {Rodr{\'\i}guez-Su{\'a}rez}, \citenamefont
  {Azevedo},\ and\ \citenamefont {Rezende}}]{r19}%
  \BibitemOpen
  \bibfield  {author} {\bibinfo {author} {\bibfnamefont {J.}~\bibnamefont
  {Holanda}}, \bibinfo {author} {\bibfnamefont {O.~A.}\ \bibnamefont {Santos}},
  \bibinfo {author} {\bibfnamefont {R.}~\bibnamefont {Cunha}}, \bibinfo
  {author} {\bibfnamefont {J.}~\bibnamefont {Mendes}}, \bibinfo {author}
  {\bibfnamefont {R.}~\bibnamefont {Rodr{\'\i}guez-Su{\'a}rez}}, \bibinfo
  {author} {\bibfnamefont {A.}~\bibnamefont {Azevedo}}, \ and\ \bibinfo
  {author} {\bibfnamefont {S.}~\bibnamefont {Rezende}},\ }\bibfield  {title}
  {\enquote {\bibinfo {title} {Longitudinal spin seebeck effect in permalloy
  separated from the anomalous nernst effect: Theory and experiment},}\
  }\href@noop {} {\bibfield  {journal} {\bibinfo  {journal} {Physical Review
  B}\ }\textbf {\bibinfo {volume} {95}},\ \bibinfo {pages} {214421} (\bibinfo
  {year} {2017})}\BibitemShut {NoStop}%
\bibitem [{\citenamefont {Slachter}\ \emph {et~al.}(2010)\citenamefont
  {Slachter}, \citenamefont {Bakker}, \citenamefont {Adam},\ and\ \citenamefont
  {van Wees}}]{r24}%
  \BibitemOpen
  \bibfield  {author} {\bibinfo {author} {\bibfnamefont {A.}~\bibnamefont
  {Slachter}}, \bibinfo {author} {\bibfnamefont {F.~L.}\ \bibnamefont
  {Bakker}}, \bibinfo {author} {\bibfnamefont {J.-P.}\ \bibnamefont {Adam}}, \
  and\ \bibinfo {author} {\bibfnamefont {B.~J.}\ \bibnamefont {van Wees}},\
  }\bibfield  {title} {\enquote {\bibinfo {title} {Thermally driven spin
  injection from a ferromagnet into a non-magnetic metal},}\ }\href@noop {}
  {\bibfield  {journal} {\bibinfo  {journal} {Nature Physics}\ }\textbf
  {\bibinfo {volume} {6}},\ \bibinfo {pages} {879} (\bibinfo {year}
  {2010})}\BibitemShut {NoStop}%
\bibitem [{\citenamefont {Brataas}, \citenamefont {Nazarov},\ and\
  \citenamefont {Bauer}(2001)}]{rr30}%
  \BibitemOpen
  \bibfield  {author} {\bibinfo {author} {\bibfnamefont {A.}~\bibnamefont
  {Brataas}}, \bibinfo {author} {\bibfnamefont {Y.~V.}\ \bibnamefont
  {Nazarov}}, \ and\ \bibinfo {author} {\bibfnamefont {G.~E.}\ \bibnamefont
  {Bauer}},\ }\bibfield  {title} {\enquote {\bibinfo {title} {Spin-transport in
  multi-terminal normal metal-ferromagnet systems with non-collinear
  magnetizations},}\ }\href@noop {} {\bibfield  {journal} {\bibinfo  {journal}
  {The European Physical Journal B-Condensed Matter and Complex Systems}\
  }\textbf {\bibinfo {volume} {22}},\ \bibinfo {pages} {99--110} (\bibinfo
  {year} {2001})}\BibitemShut {NoStop}%
\bibitem [{\citenamefont {Kimura}, \citenamefont {Hamrle},\ and\ \citenamefont
  {Otani}(2005)}]{r29}%
  \BibitemOpen
  \bibfield  {author} {\bibinfo {author} {\bibfnamefont {T.}~\bibnamefont
  {Kimura}}, \bibinfo {author} {\bibfnamefont {J.}~\bibnamefont {Hamrle}}, \
  and\ \bibinfo {author} {\bibfnamefont {Y.}~\bibnamefont {Otani}},\ }\bibfield
   {title} {\enquote {\bibinfo {title} {Estimation of spin-diffusion length
  from the magnitude of spin-current absorption: Multiterminal
  ferromagnetic/nonferromagnetic hybrid structures},}\ }\href@noop {}
  {\bibfield  {journal} {\bibinfo  {journal} {Physical Review B}\ }\textbf
  {\bibinfo {volume} {72}},\ \bibinfo {pages} {014461} (\bibinfo {year}
  {2005})}\BibitemShut {NoStop}%
\bibitem [{\citenamefont {Toka{\c{c}}}\ \emph {et~al.}(2015)\citenamefont
  {Toka{\c{c}}}, \citenamefont {Bunyaev}, \citenamefont {Kakazei},
  \citenamefont {Schmool}, \citenamefont {Atkinson},\ and\ \citenamefont
  {Hindmarch}}]{r21}%
  \BibitemOpen
  \bibfield  {author} {\bibinfo {author} {\bibfnamefont {M.}~\bibnamefont
  {Toka{\c{c}}}}, \bibinfo {author} {\bibfnamefont {S.}~\bibnamefont
  {Bunyaev}}, \bibinfo {author} {\bibfnamefont {G.}~\bibnamefont {Kakazei}},
  \bibinfo {author} {\bibfnamefont {D.}~\bibnamefont {Schmool}}, \bibinfo
  {author} {\bibfnamefont {D.}~\bibnamefont {Atkinson}}, \ and\ \bibinfo
  {author} {\bibfnamefont {A.}~\bibnamefont {Hindmarch}},\ }\bibfield  {title}
  {\enquote {\bibinfo {title} {Interfacial structure dependent spin mixing
  conductance in cobalt thin films},}\ }\href@noop {} {\bibfield  {journal}
  {\bibinfo  {journal} {Physical review letters}\ }\textbf {\bibinfo {volume}
  {115}},\ \bibinfo {pages} {056601} (\bibinfo {year} {2015})}\BibitemShut
  {NoStop}%
\bibitem [{\citenamefont {Vlaminck}\ and\ \citenamefont
  {Bailleul}(2008)}]{r27}%
  \BibitemOpen
  \bibfield  {author} {\bibinfo {author} {\bibfnamefont {V.}~\bibnamefont
  {Vlaminck}}\ and\ \bibinfo {author} {\bibfnamefont {M.}~\bibnamefont
  {Bailleul}},\ }\bibfield  {title} {\enquote {\bibinfo {title}
  {Current-induced spin-wave doppler shift},}\ }\href@noop {} {\bibfield
  {journal} {\bibinfo  {journal} {Science}\ }\textbf {\bibinfo {volume}
  {322}},\ \bibinfo {pages} {410--413} (\bibinfo {year} {2008})}\BibitemShut
  {NoStop}%
\bibitem [{\citenamefont {Sola}\ \emph {et~al.}(2017)\citenamefont {Sola},
  \citenamefont {Bougiatioti}, \citenamefont {Kuepferling}, \citenamefont
  {Meier}, \citenamefont {Reiss}, \citenamefont {Pasquale}, \citenamefont
  {Kuschel},\ and\ \citenamefont {Basso}}]{r31}%
  \BibitemOpen
  \bibfield  {author} {\bibinfo {author} {\bibfnamefont {A.}~\bibnamefont
  {Sola}}, \bibinfo {author} {\bibfnamefont {P.}~\bibnamefont {Bougiatioti}},
  \bibinfo {author} {\bibfnamefont {M.}~\bibnamefont {Kuepferling}}, \bibinfo
  {author} {\bibfnamefont {D.}~\bibnamefont {Meier}}, \bibinfo {author}
  {\bibfnamefont {G.}~\bibnamefont {Reiss}}, \bibinfo {author} {\bibfnamefont
  {M.}~\bibnamefont {Pasquale}}, \bibinfo {author} {\bibfnamefont
  {T.}~\bibnamefont {Kuschel}}, \ and\ \bibinfo {author} {\bibfnamefont
  {V.}~\bibnamefont {Basso}},\ }\bibfield  {title} {\enquote {\bibinfo {title}
  {Longitudinal spin seebeck coefficient: heat flux vs. temperature difference
  method},}\ }\href@noop {} {\bibfield  {journal} {\bibinfo  {journal}
  {Scientific reports}\ }\textbf {\bibinfo {volume} {7}},\ \bibinfo {pages}
  {46752} (\bibinfo {year} {2017})}\BibitemShut {NoStop}%
\bibitem [{\citenamefont {Chuang}\ \emph {et~al.}(2017)\citenamefont {Chuang},
  \citenamefont {Su}, \citenamefont {Wu},\ and\ \citenamefont {Huang}}]{r32}%
  \BibitemOpen
  \bibfield  {author} {\bibinfo {author} {\bibfnamefont {T.-C.}\ \bibnamefont
  {Chuang}}, \bibinfo {author} {\bibfnamefont {P.}~\bibnamefont {Su}}, \bibinfo
  {author} {\bibfnamefont {P.}~\bibnamefont {Wu}}, \ and\ \bibinfo {author}
  {\bibfnamefont {S.~Y.}\ \bibnamefont {Huang}},\ }\bibfield  {title} {\enquote
  {\bibinfo {title} {Enhancement of the anomalous nernst effect in
  ferromagnetic thin films},}\ }\href@noop {} {\bibfield  {journal} {\bibinfo
  {journal} {Physical Review B}\ }\textbf {\bibinfo {volume} {96}},\ \bibinfo
  {pages} {174406} (\bibinfo {year} {2017})}\BibitemShut {NoStop}%
\bibitem [{\citenamefont {Uchida}\ \emph {et~al.}(2010)\citenamefont {Uchida},
  \citenamefont {Adachi}, \citenamefont {Ota}, \citenamefont {Nakayama},
  \citenamefont {Maekawa},\ and\ \citenamefont {Saitoh}}]{r22}%
  \BibitemOpen
  \bibfield  {author} {\bibinfo {author} {\bibfnamefont {K.-i.}\ \bibnamefont
  {Uchida}}, \bibinfo {author} {\bibfnamefont {H.}~\bibnamefont {Adachi}},
  \bibinfo {author} {\bibfnamefont {T.}~\bibnamefont {Ota}}, \bibinfo {author}
  {\bibfnamefont {H.}~\bibnamefont {Nakayama}}, \bibinfo {author}
  {\bibfnamefont {S.}~\bibnamefont {Maekawa}}, \ and\ \bibinfo {author}
  {\bibfnamefont {E.}~\bibnamefont {Saitoh}},\ }\bibfield  {title} {\enquote
  {\bibinfo {title} {Observation of longitudinal spin-seebeck effect in
  magnetic insulators},}\ }\href@noop {} {\bibfield  {journal} {\bibinfo
  {journal} {Applied Physics Letters}\ }\textbf {\bibinfo {volume} {97}},\
  \bibinfo {pages} {172505} (\bibinfo {year} {2010})}\BibitemShut {NoStop}%
\bibitem [{\citenamefont {Kannan}\ \emph {et~al.}(2017)\citenamefont {Kannan},
  \citenamefont {Fan}, \citenamefont {Celik}, \citenamefont {Han},\ and\
  \citenamefont {Xiao}}]{r23}%
  \BibitemOpen
  \bibfield  {author} {\bibinfo {author} {\bibfnamefont {H.}~\bibnamefont
  {Kannan}}, \bibinfo {author} {\bibfnamefont {X.}~\bibnamefont {Fan}},
  \bibinfo {author} {\bibfnamefont {H.}~\bibnamefont {Celik}}, \bibinfo
  {author} {\bibfnamefont {X.}~\bibnamefont {Han}}, \ and\ \bibinfo {author}
  {\bibfnamefont {J.~Q.}\ \bibnamefont {Xiao}},\ }\bibfield  {title} {\enquote
  {\bibinfo {title} {Thickness dependence of anomalous nernst coefficient and
  longitudinal spin seebeck effect in ferromagnetic ni x fe 100- x films},}\
  }\href@noop {} {\bibfield  {journal} {\bibinfo  {journal} {Scientific
  reports}\ }\textbf {\bibinfo {volume} {7}},\ \bibinfo {pages} {6175}
  (\bibinfo {year} {2017})}\BibitemShut {NoStop}%
\end{thebibliography}%


%
\end{document}